\documentclass[10pt,conference]{IEEEtran}
\usepackage{cite}
\usepackage{amsmath,amssymb,amsfonts}
\usepackage{algorithmic}
\usepackage{graphicx}
\usepackage{textcomp}
\usepackage{xcolor}
\def\BibTeX{{\rm B\kern-.05em{\sc i\kern-.025em b}\kern-.08em
T\kern-.1667em\lower.7ex\hbox{E}\kern-.125emX}}

\usepackage[hyphens]{url}
\usepackage{hyperref}
\usepackage[hyphenbreaks]{breakurl}

\begin{document}

%\title{A research agenda for automated detection of flakiness in quantum software bug reports}
\title{Automated flakiness detection in quantum software bug reports}

%% used to denote sha qred contribution to the research.
% \author{Lei Zhang}
% \affiliation{%
%   \institution{University of Maryland, Baltimore County}
%   \city{Baltimore}
%   \state{Maryland}
% \email{leizhang@umbc.edu}

% \author{Andriy Miranskyy}
% \affiliation{%
%   \institution{Toronto Metropolitan University}
%   \city{Toronto}
%   \country{Canada}}
% \email{avm@torontomu.ca}
\author{\IEEEauthorblockN{Lei Zhang}
\IEEEauthorblockA{\textit{Department of Information Systems} \\
\textit{University of Maryland, Baltimore County}\\
Maryland, USA \\
leizhang@umbc.edu}
\and
\IEEEauthorblockN{Andriy Miranskyy}
\IEEEauthorblockA{\textit{Department of Computer Science} \\
\textit{Toronto Metropolitan University}\\
Ontario, Canada \\
avm@torontomu.ca}
}
%\renewcommand{\shortauthors}{Zhang and Miranskyy}

\maketitle
\begin{abstract}
%In the evolving landscape of quantum software engineering, both researchers and practitioners are pursuing novel methods and tools to advance quantum software development. There are multiple factors driving this trend, e.g., the relatively short history of quantum software engineering. One of the most critical factors is the distinctive natures of quantum mechanics, such as indeterminacy. This indeterminate nature poses a significant challenge to the testing and debugging process in quantum software development. 

A flaky test yields inconsistent results upon repetition, posing a significant challenge to software developers. An extensive study of their presence and characteristics has been done in classical computer software but not quantum computer software. In this paper, we outline challenges and potential solutions for the automated detection of flaky tests in bug reports of quantum software. We aim to raise awareness of flakiness in quantum software and encourage the software engineering community to work collaboratively to solve this emerging challenge.

%In classical software engineering, flaky tests characterized by unpredictable and random behaviors are well-known and well-studied. This prompts an interesting research question in the quantum world: Does the inherent randomness in quantum programs lead to flaky tests? Our previous study reveals that flakiness is a pervasive issue in quantum software. Building upon this foundation, our next research question is how to identify and mitigate flakiness in quantum programs automatically. Our ultimate goal is to improve the testing and debugging of quantum programs by addressing the challenges posed by the stochastic nature of quantum programs and pave the way for more efficient and reliable quantum software development.
\end{abstract}

% \begin{IEEEkeywords}
% flaky tests, quantum software testing, quantum software engineering
% \end{IEEEkeywords}

\section{Introduction}\label{sec:intro}

Tests run on the same code sometimes produce different results, e.g., passing sometimes, failing other times. Such tests are called flaky tests. Developers of programs for classical computers (CCs) are plagued by these misleading signals provided by flaky tests. At Google, in 2014, 73K out of 1.6M (4.56\%) of test failures were caused by flaky tests~\cite{luo2014empirical}; in 2017, 1.5\% of 4.2M tests were flaky~\cite{micco2017state, memon2017taming}. In the literature, flaky tests in CC programs have been well studied in terms of empirical analysis~\cite{luo2014empirical, gruber2021empirical, lam2020study, lam2019root, parry2021survey}, automated detection~\cite{ziftci2020flake, bell2018deflaker, lam2019idflakies, dutta2020detecting}, and mitigation techniques~\cite{dutta2021flex, wang2022ipflakies, fatima2022flakify}. 

The evolution of quantum computing has led to the growth and expansion of quantum software development. However, more efforts are needed to establish quantum software engineering best practices, novel methods, and tools to address the challenges brought by quantum computing~\cite{miranskyy2019testing, miransky2020bug, miranskyy2021testing, zhao2020quantum}. For instance, one of the challenges is to analyze flaky tests in quantum software. %{ICSE'19, ICSE'20, Zhao’20 or similar review}.

%As part of this effort, the question arises: how prevalent are flaky tests in programs for quantum computers (QCs), and do they differ from flaky tests for CCs?

Conceptually, we know that a flaky test can be caused by two factors: random behaviors in the source code and problematic tests~\cite{luo2014empirical,eck2019understanding,parry2021survey,gruber2021empirical,dutta2020detecting,dutta2021flex}. Quantum programs are inherently non-deterministic. The randomness comes from a variety of sources. For example, it can be caused by the physical properties of quantum systems (e.g., quantum indeterminacy~\cite[Ch. 1]{marinescu2011classical}) or hardware issues (e.g., measurement errors or networking problems) or both (e.g., quantum decoherence~\cite[Ch. 7]{nielsen_chuang_2010}). In a simulation of a QC on a CC, pseudo-random number generators (PRNGs) are used to emulate these sources of randomness. Randomness from all these sources leads to a distribution of output values, which may result in flaky tests.

As part of this effort, the question arises: how prevalent are flaky tests in programs for quantum computers (QCs), and do they differ from flaky tests for CCs?
To empirically assess the effect of flaky tests on QC programs, we explored the code and bug-tracking repositories of 14 quantum software~\cite{zhang_identifying_2023}. We estimate that flaky tests account for 0.26\% to 1.85\% of bug reports in 12 software packages (46 unique flaky test reports in total).\footnote{The dataset is available at \url{https://doi.org/10.5281/zenodo.7888639}.}

In these test reports, we identify and categorize eight types of flakiness and seven common fixes. While flakiness types for CC and QC are similar, the most common causes and fixes are not. Randomness is the most common cause of flakiness in QC, and the most common solution is to fix PRNG seeds. The findings for CC vary depending on programming language platforms, e.g., test order dependency takes the top place of root causes for Python program~\cite{luo2014empirical}. Moreover, quantum programmers do not use some recent countermeasures (e.g., ~\cite{dutta2020detecting,dutta2021flex}) developed by software engineers to deal with flaky tests. 

Finding the root causes and fixes of flaky tests will become more challenging as quantum hardware develops and quantum software becomes larger and more complex. Thus, it would be helpful\footnote{Classification of their root causes and fix patterns, as well as mitigation and remediation, should also be considered in the future. As we focus on detection in this paper, we intentionally leave out these topics. 
However, it is important to note that some of the existing research for addressing flakiness in software for CC---such as Oracle approximation~\cite{nejadgholi2019study} or fuzz testing tools~\cite{dutta2018testing}---can be applied to QC software. } to automate the detection of flaky tests in a bug report. The following are some possible approaches and challenges to achieving this objective.

\section{Method and Challenges}\label{sec:method}
%Detecting flaky tests and identifying their root causes and fixing patterns can be framed as a supervised or semi-supervised learning problem. The approaches and challenges in this section are grouped by the steps in a typical machine learning (ML) pipeline. 
Automated detection of flaky tests can be framed as a supervised or semi-supervised learning problem. The approaches and challenges in this section are grouped by the steps in a typical machine learning (ML) pipeline. 

% With the development of quantum hardware, quantum software will also increase in terms of size and complexity. Thus, quantum software developers must utilize automated methods and tools to improve their productivity and software quality. Our ultimate goal of this project is to automatically detect flaky tests from existing bug reports in quantum programs. To achieve this goal, we propose to four steps as follows.\footnote{For future work, we will also be interested in mitigation of flakiness in terms of automated fixing techniques.} %We propose to conduct a two-phase research project to achieve this goal. In the first phase, we propose to adopt ML techniques to detect flaky tests from artificial datasets. In the second phase, we aim to use natural language processing techniques to extract information from real-world bug reports and feed the extracted information to the ML models (in the first step) for flaky test prediction.  In this paper, we focus on the first phase, and we propose a research agenda for it as follows. 

\textit{Step 1: Dataset preparation.} This step involves preparing a dataset to train and test ML models. Each observation in the dataset is mapped to a specific bug report. Observation features are derived from the text and metadata of the bug report. Labels on observations are binary, i.e., a bug report is related to flaky tests (positive label) or not (negative label).

At least three \textit{challenges} stem from the nature of the dataset: 46 of the 5,484 bug reports in 12 popular software relate to flaky tests~\cite{zhang_identifying_2023}. First, the amount of data is limited (due to the relatively short history of quantum computing software). We may compensate for the lack of data by enriching training data with CC software data (since the symptoms may be similar, although the underlying causes may differ).

Additionally, the dataset is highly imbalanced (which is common for CC and QC software). Standard Data Science techniques, such as oversampling, can be used to combat the imbalance.

Finally, the dataset may contain noisy labels. Since multiple co-authors vetted the observations, we are confident that the positive labels are accurate. However, negative labels may be incorrect since potential flaky tests were identified using keywords, and we cannot guarantee that the keywords set was exhaustive~\cite{zhang_identifying_2023}.

% Step 1, dataset preparation (creating features with labels). In this step, we will prepare a dataset for training and testing ML models (e.g., classification models and/or artificial neural networks). The dataset contains extracted features with labels (if it is a flaky test rather than a regular issue).  

% The challenge in this step is the limited data. The data for QSE are limited due to the relatively short history of quantum computing. Based on our previous study~\cite{zhang_identifying_2023}, we only detect 46 flaky tests from 5,484 issue reports in 12 software repositories, making the dataset highly imbalanced and relatively small. For training ML models, the size and balance of the training dataset is critical to optimize the performance of the models. For small training datasets, we can adopt simple models, e.g., random forests or support vector machines from the classical classification algorithm group.     

\textit{Step 2: Model development.} To classify the dataset labels, a model should be designed and trained. As discussed in Step 1, the \textit{challenge} arises from the nature of the dataset again. The dataset is small, so we may have to resort to classical ML models (e.g., Random Forest~\cite{breiman2001random} or XGBoost~\cite{chen2016xgboost}). However, we may also explore artificial neural networks, particularly the transfer learning approach (training the initial model on the larger CC dataset). We can mitigate dataset imbalance during training using standard techniques, such as weight balancing. We can also leverage semi-supervised learning techniques by treating some noisy negative labels as unlabelled.

%Step 2, model development. We will develop the model in Python, and we will train the models with the dataset prepared in Step 1. We propose to try both classical ML models and artificial neural networks (ANNs). For ANNs, developers can choose semi-supervised learning approaches if the dataset contains unlabeled flaky tests. In semi-supervised learning, ANNs will be trained on a small set of labeled data and produce pseudo-labels. The true labeled data and pseudo-labeled data will be combined to train and improve the model.

\textit{Step 3: Model validation.} The performance of trained models should be evaluated. As the dataset contains noisy labels, validation of model performance can be \textit{challenging}. Active learning techniques (see~\cite{ren2021survey} for review) may be used to identify and fix potentially mislabeled instances in the dataset. Thus, we would be able to simultaneously improve the dataset's quality and evaluate the model's performance.

\textit{Step 4: Model deployment.} It is possible to integrate models that identify flaky bug reports into bug trackers (e.g., GitHub has a continuous integration/workflow automation feature called GitHub Actions\footnote{\url{https://github.com/features/actions}}). The \textit{challenge} here is that the dataset used to train the model may eventually become obsolete (e.g., since developers may describe flaky tests differently in future reports). It is, therefore, essential to incorporate mitigation techniques, such as periodic dataset refreshes and a feedback mechanism for detecting data and model drift.

\section{Conclusions}\label{sec:conclusions}

In our previous work, we confirmed the existence of flaky tests in quantum programs and analyzed their common causes and fixes. This paper proposes a research agenda for automatically detecting bug reports (related to flaky tests) in quantum programs. Moreover, we discuss possible approaches and challenges for producing this automation. Following the development of these detection techniques, we will investigate how flakiness risks can be mitigated and remedied in quantum software. %This paper aims to raise awareness of flakiness in quantum software within the software engineering community. We encourage community members to contribute to assessing, mitigating, and remediating flakiness risks.

% Our previous effort confirmed the prevalence of flakiness and analyzed the common causes and fixes for flakiness in quantum programs. In this paper, we propose a novel research agenda for the automated detection of flaky tests in quantum programs. The research agenda illustrates our methods to answer the research question step-by-step: We will conduct feature engineering, develop ML models, and validate the models. In addition, we also provide our insights into the research challenges and opportunities for detecting and mitigating flakiness in quantum software. We hope this paper brings awareness of quantum flakiness to the software engineering community, and we would like to see more researchers join the effort to mitigate the risk of flakiness.

%\bibliographystyle{ACM-Reference-Format-Abbr}
\bibliographystyle{IEEEtran}
\bibliography{references}

\end{document}